\documentclass[prd,eqsecnum,preprint,showpacs]{revtex4}

\usepackage{pstricks, pst-node, pst-text, pst-3d,pst-coil, epsfig,rotating}
\usepackage{amsmath,graphicx}

\global\arraycolsep=2pt
\newcommand{\ben}{\begin{equation*}}
\newcommand{\een}{\end{equation*}}
\newcommand{\bean}{\begin{eqnarray*}}
\newcommand{\eean}{\end{eqnarray*}}
\newcommand{\bnabla}{\mbox{\boldmath{$\nabla$}}}
\newcommand{\bGamma}{\mbox{\boldmath{$\Gamma$}}}
\newcommand{\bPhi}{\mbox{\boldmath{$\Phi$}}}
\newcommand{\nn}{\nonumber}
\newcommand{\be}{\begin{equation}}
\newcommand{\ee}{\end{equation}}
\newcommand{\bea}{\begin{eqnarray}}
\newcommand{\eea}{\end{eqnarray}}

\begin{document} 

\title{Multiple Scattering Casimir Force Calculations:
Layered and Corrugated Materials, Wedges, and Casimir-Polder Forces}

\author{Kimball A. Milton} 
\email{milton@nhn.ou.edu}
\author{Prachi Parashar}
\email{prachi@nhn.ou.edu}
\author{Jef Wagner} 
\email{wagner@nhn.ou.edu}
\affiliation{Oklahoma Center for High Energy Physics, and
H. L. Dodge Department of Physics and Astronomy, University of Oklahoma,
Norman, OK 73019-2061, USA
}

\author{I. Cavero-Pel\'aez}
\email{cavero@nhn.ou.edu}
\affiliation{Theoretical Physics Department, Zaragoza University,
Zaragoza, Spain}

\date{\today}
\begin{abstract}
Various applications of the multiple scattering technique to
calculating Casimir energy are described.  These include the
interaction between dilute bodies of various sizes and shapes,
temperature dependence, interactions with multilayered and corrugated bodies,
and new examples of exactly solvable separable bodies.
\end{abstract}
\maketitle
\section{Introduction}
For many years, calculation of Casimir or quantum vacuum energies
were confined to systems of very high symmetry, such as parallel
plates \cite{casimir,lifshitz}.  Self-energies of spheres  \cite{Boyer:1968uf}
and cylinders \cite{deraad} were also calculated.
To calculate the force between curved surfaces, such as that in the
experimentally realizable spherical lens above a plate, resort had to be
made to the ``proximity force approximation'' \cite{derjaguin},
 which, while exact for
zero separation, was subject to unknown corrections for finite distances
between the surfaces.

That situation has radically changed.  The ``multiple scattering'' technique,
not merely increase in computing power, has been responsible for much of
the improvement.  The method, related to the Krein trace formula \cite{krein},
was used early on to construct the Lifshitz formula for the quantum
fluctuation force between dielectric plates \cite{renne}.
It was the basis of the classic work by Balian and Duplantier \cite{Balian}.
But recently it was realized by many people that it could be applied
to practical calculations of Casimir forces between bodies of essentially
arbitrary shape and orientation 
\cite{Wirzba:2007bv, Bordag:2008gj, maianeto08, Emig:2007cj, rahi, reid}.  
For more complete references and background,
see Refs.~\cite{Milton:2007wz,Milton:2008st}.
Complementary developments by Gies and collaborators on the worldline
method \cite{Gies:2006cq, Gies:2006bt, Gies:2006xe, Gies:2003cv},
and direct numerical methods \cite{capasso, rodriguez, rod} should also
be cited.

In this paper we summarize the work of the Oklahoma group in applying the
multiple scattering technique to a variety of problems.  The hope is that
results will be forthcoming that would be of use in the design of 
nanotechnology.  We will merely summarize some representative results
here; for more complete details and further applications the reader is
referred to the original previous and forthcoming papers.

\section{Multiple Scattering Technique}
\label{sec:ms}
For simplicity, we first restrict attention to the quantum vacuum forces
arising from a massless scalar field. 
The multiple scattering approach starts from the well-known formula
for the vacuum energy or Casimir energy 
\be
E=\frac{i}{2\tau}\mbox{Tr}\ln G\to\frac{i}{2\tau} \mbox{Tr}\ln G G_0^{-1},
\ee
where $\tau$ is the ``infinite'' time that the
configuration exists, and 
$G$ ($G_0$) is the Green's function, which satisfies the differential
equation
\be
(-\partial^2+V)G=1, \quad -\partial^2 G_0=1, \ee
in terms of the background potential $V$.

Now we define the $T$-matrix,
\be
T=S-1=V(1+G_0V)^{-1}.\ee
If the potential has two disjoint parts,
$V=V_1+V_2$,
it is easy to derive
\be
T=(V_1+V_2)(1-G_0T_1)(1-G_0T_1G_0T_2)^{-1}(1-G_0T_2),
\label{tgtg}\ee
\be
T_i=V_i(1+G_0V_i)^{-1},\quad i=1,2.\ee
Thus, we can write the general expression for 
the interaction between the two bodies (potentials):
\be
E_{12}= -\frac{i}{2\tau}\mbox{Tr}\ln(1-G_0T_1G_0T_2)
=-\frac{i}{2\tau}\mbox{Tr}\ln(1-V_1G_1V_2G_2),\ee
where $G_i=(1+G_0V_i)^{-1}G_0,\quad i=1,2$.

\subsection{Exact Results for Weak Coupling}
In weak coupling, where the potentials are very small, 
it is possible to derive the exact (scalar) interaction
between two potentials \cite{Wagner:2008qq}, either in 2- or 3-dimensions:
\begin{subequations}
\be
2D:\quad \frac{E}{L_z}=-\frac1{32\pi^3}\int(d\mathbf{r_\perp})
(d\mathbf{r_\perp'})
\frac{V_1(\mathbf{r_\perp})V_2(\mathbf{r_\perp'})}{|\mathbf{r-r'}|^2},
\label{2d}
\ee
where $L_z$ is the ``infinite'' length in the translationally-invariant
direction, and
\be 3D: \quad E=-\frac1{64\pi^3}\int(d\mathbf{r})(d\mathbf{r'})
\frac{V_1(\mathbf{r})V_2(\mathbf{r'})}{|\mathbf{r-r'}|^3}.\ee
\end{subequations}

\begin{figure}[t]
\centering
\includegraphics{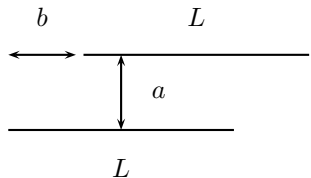}
\caption{\label{figoffpl} Two ribbons of finite width, offset from each other.
The ribbons have infinite extent out of the page.}
\end{figure}

Consider two plates (ribbons) of finite width $L$, offset by an amount $b$,
separated by a distance $a$, as shown in Fig.~\ref{figoffpl},
\begin{subequations}
\bea
V_1(\mathbf{r_\perp})&=&\lambda_1\delta(y)\theta(x)\theta(L-x),\\
V_2(\mathbf{r_\perp'})&=&\lambda_2\delta(y'-a)\theta(x'-b)\theta(L+b-x'),
\eea
\end{subequations}
Equation (\ref{2d}) gives an explicit result for the energy 
of interaction between the plates:
\be \frac{E}{L_z}=-\frac{\lambda_1\lambda_2}{32\pi^3}\left[-2g(b/a)
+g((L-b)/a)+g((L+b)/a)\right],\ee
where 
\be
g(x)=x \tan^{-1}x-\frac12\ln(1+x^2)=-\mbox{Re}(1+ix)\ln(1+ix).\ee
We can consider arbitrary lengths and orientations, in 3 dimensions, for
the plates \cite{Wagner:2008qq}.

\begin{figure}
\centering
\includegraphics{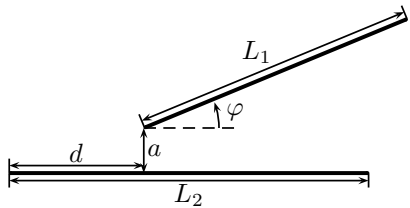}
\caption{\label{fig2} Two ribbons of finite width tilted with respect
to each other.}
\end{figure}

For example, we can consider tilted plates, as shown in Fig.~\ref{fig2}.
Explicit interaction energies can be given in terms of $\mbox{Ti}_2$,
the inverse tangent integral. For fixed distance $D$ between the center
of masses of the plates, and for $L_1\to L$, $L_2\to\infty$,
$d\to-\infty$ (that is, the upper plate is completely above the
much wider lower plate), and if $D>\frac{L}2$, 
the equilibrium position of the upper plate is at $\phi=\pi/2$.  That
means there is a torque on
the upper plate tending to  orient it to be  perpendicular to the lower plate.

\begin{figure}
\centering
\includegraphics{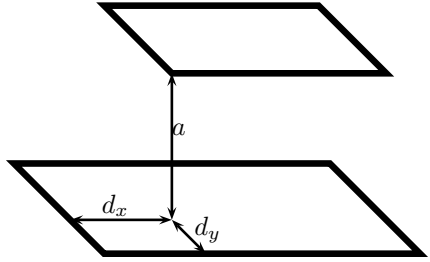}
\caption{\label{fig3} Two parallel finite rectangular plates.}
\end{figure}

\begin{figure}
\centering
\includegraphics{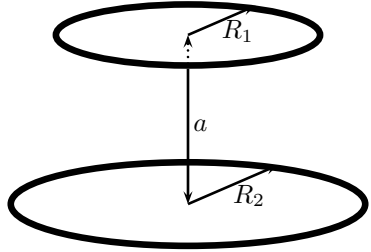}
\caption{\label{fig4} Coaxial disks}
\end{figure}

We can also examine the interaction between rectangular parallel plates
of finite area, as illustrated in Fig.~\ref{fig3}.
 As the perpendicular separation between the plates $a$ tends to $0$,
the force per area on the plates has the expansion 
\be
\frac{F}A=-\frac{\lambda_1\lambda_2}{32\pi^2a^2}(1+c_1a+c_2a^2+\dots).
\ee
The leading correction to the Lifshitz formula, $c_1$, has
a simple geometrical interpretation..
If the upper plate is completely above the lower plate, the leading
correction vanishes, $c_1=0$.  On the other hand, 
if the plates are of the same size and aligned,  the correction is
geometrical:
\be
c_1=-\frac1\pi\frac{\mbox{Perimeter}}{\mbox{Area}}.\label{geometry}
\ee
In a similar vein, we can consider coaxial disks, as shown in
Fig.~\ref{fig4}.
  In this case, the leading correction is completely
consistent with the result for rectangular plates, that is, the leading
correction vanishes if the disks are of unequal size, $R_1<R_2$, $c_1=0$, 
while if the are equal, $R_1=R_2$, the correction is again given
by Eq.~(\ref{geometry}).

\begin{figure}
\centering
\includegraphics{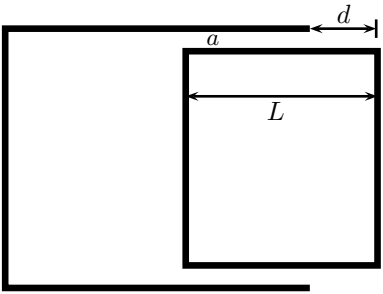}
\caption{\label{fig5} The Casimir pistol. Is it possible to
choose parameters so that the ``bullet'' is expelled from the pistol
due to quantum vacuum forces?}
\end{figure}

We can summarize the salient features for two thin plates as follows:
 Two plates of finite size experience a lateral force so that
they wish to align in the position of maximum symmetry.
 In this symmetric configuration, there is a torque about the center of mass 
of a smaller plate so
that it tends to seek perpendicular orientation with respect to the
larger plate.  Finally, the first
short-distance  correction to the normal Lifshitz force is geometrical.
The results are of relevance to the recent discussions of the Casimir pistol
\cite{fulling}, shown in Fig.~\ref{fig5}.

\section{Summing van der Waals forces}
In the preceding section the focus was on scalar field theory.
More relevant is electrodynamics, and the quantum vacuum forces between
conducting and dielectric bodies.  Weak-coupling in this regime
is described by the Casimir-Polder force between molecules 
\cite{Casmir:1947hx}.  
The following discussion is based on Ref.~\cite{Milton:2008vr}.
(For a summary of very extensive work in the nonretarded regime, see
Ref.~\cite{parsegian}.) 

The (retarded dispersion) van der Waals (vdW) potential between polarizable
molecules is given by
\be
V=-\frac{23}{4\pi}\frac{\alpha_1\alpha_2}{r^7},\quad \alpha=\frac{\varepsilon
-1}{4\pi N}.\ee
This allows us to consider in the same vein
as in Sec.~\ref{sec:ms} the electromagnetic
interactions between distinct dilute dielectric bodies of arbitrary shape.
This vdW potential may be directly derived from the action
\be
W=-\frac{i}2 \mbox{Tr}\,\ln\bGamma\bGamma_0^{-1}\approx- \frac{i}2
\mbox{Tr}\,V_1\bGamma_0 V_2\bGamma_0,\quad V_i\ll1,\label{trln}\ee
where $V_i=\varepsilon_i-1$ and
\be
\bGamma_0=
\bnabla\times\bnabla\times\mathbf{1}\frac{e^{-|\zeta||\mathbf{r-r'}|}}
{4\pi|\mathbf{r-r'}|}-\mathbf{1}
=(\bnabla\bnabla-\mathbf{1}\zeta^2)G_0(\mathbf{r-r'}).\ee

\subsection{Interaction between $\varepsilon$, $\mu$ bodies}
More generally, we can
consider material bodies characterized by a permittivity 
$\varepsilon(\mathbf{r})$ and a permeability $\mu(\mathbf{r})$,
so we have corresponding electric and magnetic potentials
\be
V_e(\mathbf{r})=\varepsilon(\mathbf{r})-1,\quad
V_m(\mathbf{r})=\mu(\mathbf{r})-1.
\ee
Then the trace-log in Eq.~(\ref{trln})
is ($\bPhi_0=-\frac1\zeta\bnabla\times\bGamma_0$)
\bea
\mbox{Tr}\,\ln \bGamma\bGamma_0^{-1}&=&-\mbox{Tr}\,
\ln(\mathbf{1}-\bGamma_0 V_e)-\mbox{Tr}\,
\ln(\mathbf{1}-\bGamma_0 V_m)\nonumber\\
&&\mbox{}-\mbox{Tr}\,\ln(\mathbf{1}+\bPhi_0 \mathbf{T}_e\bPhi_0 \mathbf{T}_m),
\eea
in terms of the $\mathbf{T}$-matrix,
\be
\mathbf{T}_{e,m}=V_{e,m}(\mathbf{1}-\bGamma_0 V_{e,m})^{-1}.\ee

If we have {\em disjoint\/} electric bodies, the interaction term
factorizes just as in Eq.~(\ref{tgtg}):
\be
\mbox{Tr}\,\ln \left(\mathbf{1}-\bGamma_0(V_1+V_2)\right)=-\mbox{Tr}\,\ln
(\mathbf{1}+\bGamma_0\mathbf{T}_1)
-\mbox{Tr}\,\ln
(\mathbf{1}+\bGamma_0\mathbf{T}_2)
-\mbox{Tr}\,\ln(\mathbf{1}-\bGamma_0\mathbf{T}_1
\bGamma_0\mathbf{T}_2),
\ee
so only the latter term contributes to the  interaction energy,
\be
E_{\rm int}=\frac{i}2\mbox{Tr}\ln(\mathbf{1}-\bGamma_0 \mathbf{T}_1
\bGamma_0 \mathbf{T}_2).
\ee
A similar result holds if one body is electric and the other magnetic,
\be
E_{\rm int}=-\frac{i}2\mbox{Tr}\ln(1+\bPhi_0 \mathbf{T}_1^e\bPhi_0
\mathbf{T}_2^m).\ee
Using this, it is easy to show that the Lifshitz energy between  parallel
dielectric and diamagnetic  slabs separated by a distance $a$ is
\be
E_{\varepsilon\mu}=\frac1{16\pi^3}\int d\zeta\int d^2k\bigg[\ln\left(
1-r_1 r_2'e^{-2\kappa a}\right)
+\ln\left(1-r'_1 r_2e^{-2\kappa a}\right)\bigg]\ee
where
\be
r_i=\frac{\kappa-\kappa_i}{\kappa+\kappa_i},\quad r_i'=
\frac{\kappa-\kappa'_i}{\kappa+\kappa'_i},\ee
with
$\kappa^2=k^2+\zeta^2$, $\kappa^2_1=k^2+\varepsilon\zeta^2$, $\kappa_1'
=\kappa_1/\varepsilon$, $\kappa^2_2=k^2+\mu\zeta^2$, $\kappa_2'
=\kappa_2/\mu$.
This means that
in the perfect reflecting limit, $\varepsilon\to\infty$, $\mu\to\infty$, 
we get Boyer's repulsive result \cite{boyer},
\be
E_{\rm Boyer}=+\frac78\frac{\pi^2}{720 a^3}.
\ee

\subsection{Dilute dielectrics}
\begin{figure}[h]
\centering
\includegraphics{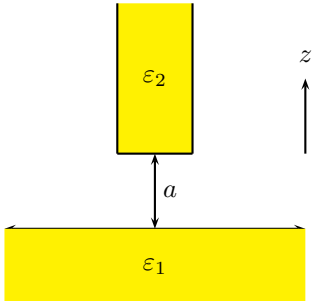}
\caption{\label{figplpl} Dielectric slab of cross section $A$ above an 
infinite dielectric plate.}
\end{figure}

We now give some exact results for dilute dielectrics, $|\varepsilon-1|\ll1$.
For example, consider the force between a semi-infinite slab and a
slab of finite cross-sectional area $A$ as shown in Fig.~\ref{figplpl}.
The force between the slabs is
\be
\frac{F}A=-\frac{23}{640\pi^2}\frac1{a^4}(\varepsilon_1-1)(\varepsilon_2-1),
\ee
which is the Lifshitz formula for infinite (dilute) slabs.
{\it Note that there is no correction due to the finite area of the
upper slab.}

We can also compute the force between a dilute dielectric sphere and 
a dilute dielectric plate, as illustrated in Fig.~{\ref{figsppl}.
\begin{figure}[h]
\centering
\includegraphics{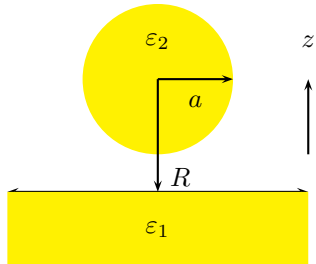}
\caption{\label{figsppl} Dilute dielectric sphere above a dilute
dielectric slab.}
\end{figure}
The energy of interaction is given by
\be
E=-\frac{23}{640\pi^2}(\varepsilon_1-1)(\varepsilon_2-1)\frac{4\pi a^3/3}{R^4}
\frac1{(1-a^2/R^2)^2},\ee
which agrees with the proximity force
approximation in the short separation limit, $R-a=\delta\ll a$:
\be
F_{\rm PFA}=2\pi a\mathcal{E}_\|(\delta)=-\frac{23}{640\pi^2}(\varepsilon_1-1)
(\varepsilon_2-1)\frac{2\pi a}{3\delta^3},\ee
with an exact correction, intermediate between that for scalar
1/2(Dirichlet+Neumann) \cite{Wirzba:2007bv,Bordag:2008gj}
and electromagnetic perfectly-conducting boundaries \cite{maianeto08}.

\subsubsection{Torque between slab and plate}
\begin{figure}
\centering
\includegraphics{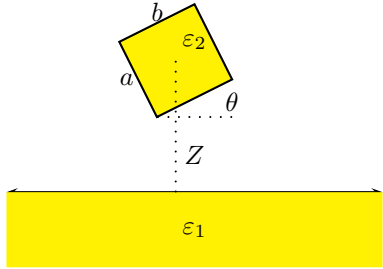}
\caption{\label{tiltedslab} Rectangular solid (with dielectric
constant $\varepsilon_2$) of side $a$, $b$, and $c$ (perpendicular
to the plane, not shown)
 a distance $Z$ above an infinite plate (with dielectric
constant $\varepsilon_1$) extending to $z=-\infty$.
The shorter side $b$ makes an angle $\theta$ with respect to the plate.}
\end{figure}
Figure \ref{tiltedslab} shows a dilute 
rectangular solid above an infinite dilute plate,
where the shorter side makes an angle $\theta$ relative to the plane.
Generically, the shorter
side wants to align with the plate, which is obvious geometrically,
since that (for fixed center of mass position) minimizes the energy.
However, if the slab has square cross section, the equilibrium
position occurs when a corner is closest to the plate, also obvious
geometrically.  But if the two sides are close enough in length,
a nontrivial equilibrium position between these extremes can
occur.  Figure \ref{equilangle} shows the equilibrium angle, for
fixed center of mass position.
\begin{figure}
\includegraphics{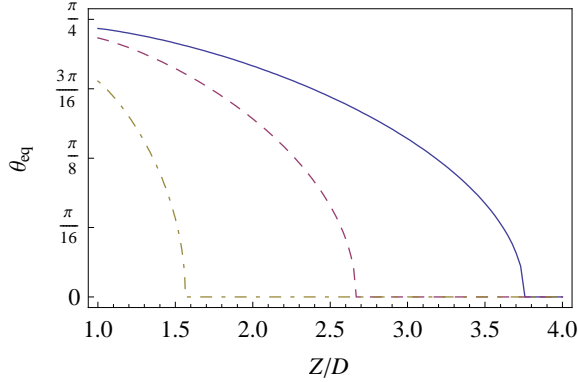}
\caption{\label{equilangle} Equilibrium angle as a function of 
separation of the body from the plane, for given $b/a$ ratios
0.95, 0.9, and 0.7, respectively given by solid, dashed, and dot-dashed
lines.}
\end{figure}
 For large enough separation, the shorter side wants to face the plate,
but for \be Z<Z_0=\frac{a}2\sqrt{\frac{2a^2+5 b^2
+\sqrt{9 a^4+20 a^2 b^2+20  b^4}}{5 \left(a^2-b^2\right)}}\ee
the equilibrium angle increases, until finally at
$Z=D=\sqrt{a^2+b^2}/2$ the slab touches the plate at an angle $\theta=
\arctan b/a$, that is, the center of mass is just above the point of
contact, about which point there is no torque.

\subsubsection{Interaction between parallel cylinders}
Two parallel dilute cylindrical bodies of radius $a$ and $b$
(of large length $L$), outside each other, with a distance $R$ between
their axes, have the interaction energy
\be
\frac{E}L=-\frac{23}{60\pi}(\varepsilon_1-1)(\varepsilon_2-1)\frac{a^2b^2}
{R^6}
\frac{1-\frac12\left(\frac{a^2+b^2}{R^2}\right)-\frac12
\left(\frac{a^2-b^2}{R^2}\right)^2}{\left[\left(1-\left(\frac{a+b}R\right)^2
\right)\left(1-\left(\frac{a-b}R\right)^2\right)\right]^{5/2}}.
\ee
This result can be analytically continued to the case when one dielectric
cylinder is entirely inside a hollowed-out cylinder within an infinite
dielectric medium \cite{Milton:2008bb}, as shown in Fig.~\ref{cylin}.
\begin{figure}
\centering
\includegraphics{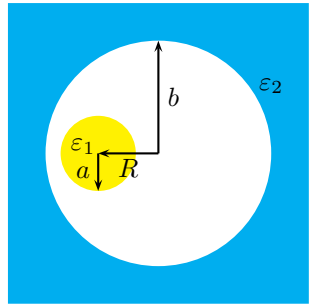}
\caption{\label{cylin} Dielectric cylinder within hollow parallel cylindrical
cavity.}
\end{figure}

\subsubsection{Interaction between spheres}
We close this section by giving the interaction energy between two
dilute spheres, of radius $a$ and $b$, respectively, separated by a distance
$R>a+b$:
\bea
E&=&-\frac{23}{1920\pi}\frac{(\varepsilon_1-1)(\varepsilon_2-1)}{R}
\Bigg\{\ln\left(\frac{1-\left(\frac{a-b}R\right)^2}{1-
\left(\frac{a+b}R\right)^2}\right)\nn\\
&&\mbox{}+\frac{4ab}{R^2}\frac{\frac{a^6-a^4b^2-a^2b^4+b^6}{R^6}-
\frac{3a^4-14a^2b^2+3b^4}{R^4}+3\frac{a^2+b^2}{R^2}-1}{\left[\left(
1-\left(\frac{a-b}R\right)^2\right)\left(1-\left(\frac{a+b}R\right)^2\right)
\right]^2}\Bigg\}.
\eea
This expression, which is rather ugly, may be verified to yield
the proximity force theorem:
\be
E\to U=-\frac{23}{640\pi}\frac{a(R-a)}{R\delta^2},\quad \delta=R-a-b\ll a,b.
\ee
It also, in the limit $b\to\infty$, $R\to\infty$ with $R-b=Z$ held
fixed, reduces to the result for the interaction of a sphere
with an infinite plate.

\section{Exact temperature results}
The scalar Casimir free energy between two weak nonoverlapping potentials
$V_1({\bf r})$ and $V_2({\bf r})$ at temperature $T$ is \cite{Milton:2009gk}
\be
E_T=-\frac{T}{32\pi^2}\int(d\mathbf{r})(d\mathbf{r'})V_1(\mathbf{r})
V_2(\mathbf{r'})\frac{\coth 2\pi T|\mathbf{r-r'}|}{|\mathbf{r-r'}|^2}.
\label{ET}
\ee
From this, we see that the free energy of interaction
 between a semitransparent plane and an arbitrarily curved
nonintersecting semitransparent surface $S$, described by the potentials
\be
V_1=\lambda_1\delta(z),\quad V_2=\lambda_2\delta(z'-z(S)),
\ee has the following form:
\be
E_T=-\frac{\lambda_1\lambda_2 T}{16\pi}\int dS\int_{2\pi T z(S)} dx
\frac{\coth x}x,
\ee
where the area integral is over the curved surface.
(The upper limit of the $x$-integral is irrelevant, since it does not
contribute to the force between the surfaces.)
This is precisely what one means by the proximity force approximation (PFA):
\be
E_{\rm PFA}=\int dS\mathcal{E}_\|(z(S)),
\ee
$\mathcal{E}_{\|}(z(S))$ being the Casimir energy between parallel
plates separated by a distance $z(S)$.
This is the theorem proved by Decca et al.~\cite{Decca:2009fg}
for the case of gravitational and Yukawa forces.

\subsection{Interaction between semitransparent spheres}
The free energy of interaction between two weakly-coupled semitransparent
spheres, described by the potentials
\be
V_1=\lambda_1\delta(r-a),\quad V_2=\lambda_2\delta(r'-b),
\ee
in terms of local spherical coordinates, the centers of which are
separated by a distance $R>a+b$, is
\bea
E_T&=&-\frac{\lambda_1\lambda_2}{16 \pi}\frac{ab}R
\bigg\{\ln\frac{1-(a-b)^2/R^2}{1-(a+b)^2/R^2}+f(2\pi T(R+a+b))\nn\\
&&\quad\mbox{}+f(2\pi T(R-a-b))-f(2\pi T(R-a+b))
-f(2\pi T(R+a-b))\bigg\},
\eea
where $f$ is given by the power series ($B_n$ is the $n$th Bernoulli 
polynomial)
\be
f(y)=\sum_{n=1}^\infty \frac{2^{2n}B_{2n}}{2n(2n-1)(2n)!}y^{2n},
\label{ltexp}
\ee
which satisfies the differential equation
\be
y\frac{d^2}{dy^2}f(y)=\coth y-\frac1y,\quad f(0)=f'(0)=0.\label{diffeq}
\ee
The differential equation may be solved numerically, yielding the results
shown in Fig.~\ref{fignum}.
\begin{figure}[tb]
  \begin{center}
  \includegraphics[width=3in]{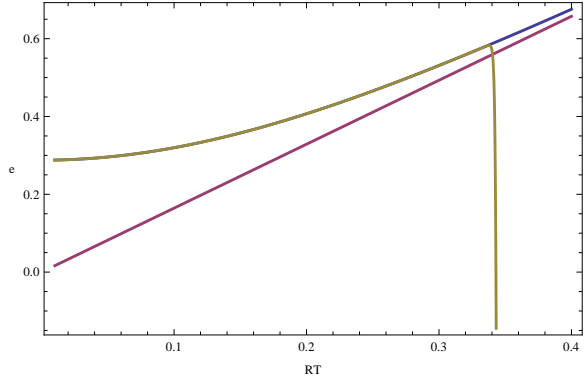}
 \caption{\label{fignum}
Finite temperature interaction energy
between two weakly-coupled semitransparent spheres, with radii $a=b=R/4$,
where $R$ is the distance between the centers of the spheres.  Shown
are the exact result, the high $T$ limit, and the truncated series expansion.
Plotted is $E=-\lambda_1\lambda_2 a b e/16\pi R$.}
  \end{center}
\end{figure}

\section{ Noncontact gears}
We consider first the interaction between corrugated planes, as shown in
Fig.~\ref{figcorru}.
\begin{figure}
\includegraphics{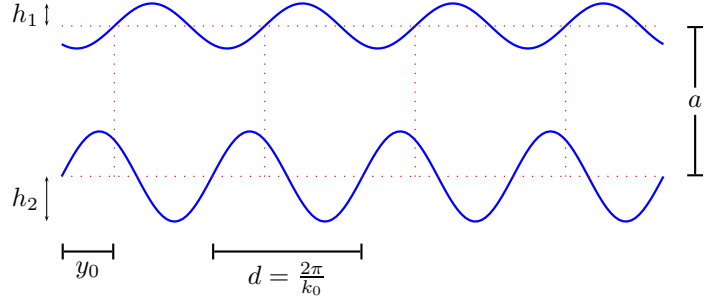}
\caption{\label{figcorru} Parallel, corrugated planes.}
\end{figure}
Here we compute the {\em lateral\/} force $F_{\rm lat}$ between the corrugated
plates, which are offset by a distance $y_0$.
The Dirichlet and electromagnetic cases were previously considered
by Emig et al.~\cite{Emig:2001dx,emig03,Buscher:2004tb}}, 
to second order in corrugation amplitudes.
We have carried out the calculations to fourth order
\cite{CaveroPelaez:2008tj}.  In weak coupling
we can calculate to all orders, and verify that fourth order is
very accurate, provided $k_0 h\ll 1$.  We express the lateral force
relative to the normal Casimir force between uncorrugated plates,
$F_{\rm Cas}^{(0)}$:
\be
\mathcal{F}=\frac{F_{\rm Lat}}{|F^{(0)}_{\rm Cas}|(h_1h_2/a^2)
 k_0a\sin(k_0y_0)}.
\ee
The weak coupling limit is shown in Fig.~\ref{wccorr}.
\begin{figure}
\centering
\epsfig{file=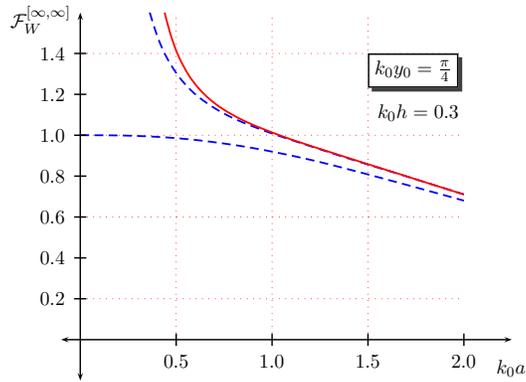, height=6cm}
\caption{\label{wccorr} Weak coupling lateral forces for corrugated planes.
Shown by the solid line is the exact lateral force, 
compared with the forces calculated in 2nd order and 4th order 
in the corrugation amplitudes.
Here we assume the two plates have equal corrugation amplitudes $h$.
The $\infty$ superscripts signify that the result is exact both in
$h$ and $a$, relative to the wavelength of the corrugations.}
\end{figure}

\subsection{Concentric corrugated cylinders}
\begin{figure}
\centering
\includegraphics{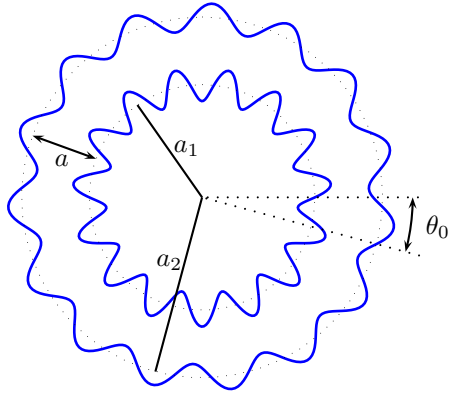}
\caption{\label{ccorr} Concentric corrugated cylinders.}
\end{figure}

\begin{figure}
\epsfig{file=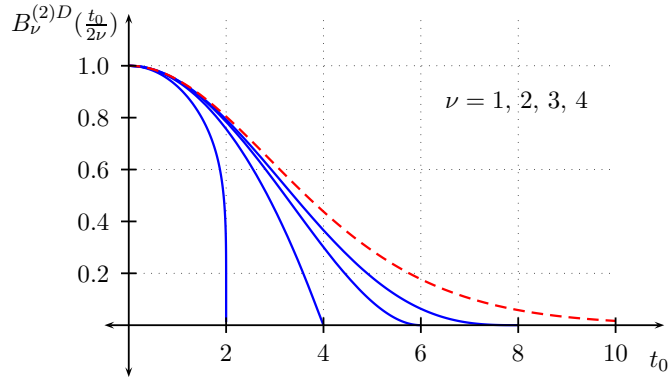,height=6cm}
\caption{\label{dcorr} Torques between corrugated cylindrical gears
calculated in 2nd order in the corrugation amplitudes, compared to the
similar result for corrugated planes.}
\end{figure}

We can also consider concentric corrugated cylinders, as shown in
Fig.~\ref{ccorr} \cite{CaveroPelaez:2008tk}.
For corrugations given by $\delta$-function potentials with
sinusoidal amplitudes:
\begin{subequations}
\bea
h_1(\theta)&=&h_1\sin\nu(\theta+\theta_0),\\
h_2(\theta)&=&h_2 \sin\nu\theta,
\eea
\end{subequations}
the torque to lowest order in the corrugations in strong coupling
(Dirichlet limit) is
\be
\frac{\tau^{(2)D}}{2\pi R L_z}=\nu\sin\nu\theta_0\frac{\pi^2}{240 a^3}
\frac{h_1}a\frac{h_2}a B_\nu^{(2)D}(\alpha), \quad \alpha=(a_2-a_1)/(a_2+a_1)
\quad R=\frac12(a_1+a_2).
\ee
Figure \ref{dcorr} shows the Dirichlet limit of
the torque on cylindrical gears.
A similar result can be found for weak coupling, which, again, has
a closed form.

\begin{figure}
\centering
\epsfig{file=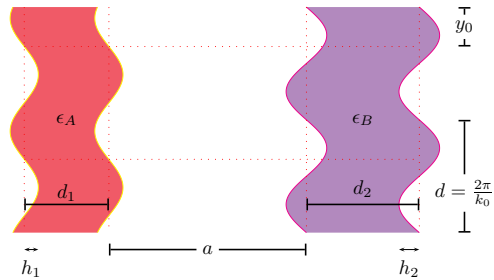,height=4cm}
\caption{\label{cds} Corrugated parallel dielectric slabs.}
\end{figure}

We are currently completing analogous calculations for dielectric
materials, illustrated by Fig.~\ref{cds}.
2nd and 4th order results should appear soon, which will complement
other recent work \cite{Lambrecht:2009zz}.

\section{Multilayered surfaces}
\begin{figure}
\epsfig{file=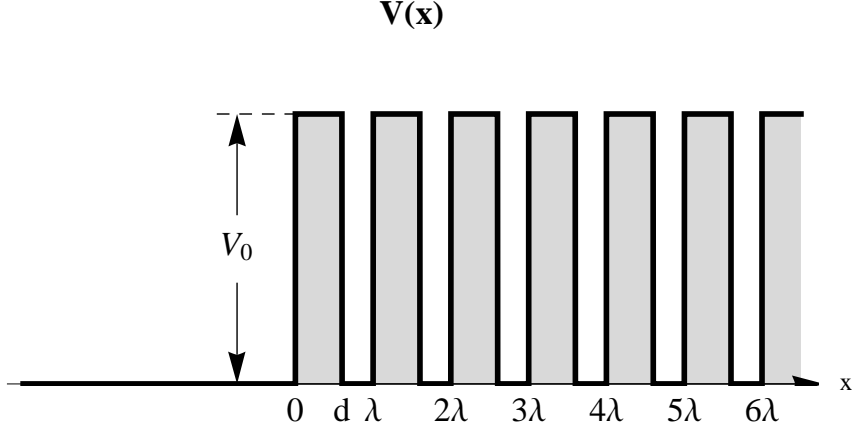,height=6cm}
\caption{\label{ml} A semi-infinite array of periodic potentials.}
\end{figure}

An example of a simple multilayered potential is given in Fig.~\ref{ml}.
To the left of this array of potentials, the reduced Green's function
has the form, in terms of 
the reflection coefficient $\mathcal{R}$ for the array:
\be
g(x,x')=\frac1{2\kappa}\left(e^{-\kappa|x-x'|}+\mathcal{R}e^{\kappa(x+x')}
\right),\quad z,z'<0.
\ee
(We can actually find the Green's function everywhere, for any piecewise
continuous potential.)

The array reflection coefficient may be readily expressed in terms
of the reflection ($R$) and transmission ($T$) 
coefficients for a single potential:
\be
\mathcal{R}=R+T e^{-\kappa a}\mathcal{R}e^{-\kappa a}(1-R e^{-2\kappa a}
\mathcal{R})^{-1},
\ee
where $a=\lambda-d$ is the distance between the potentials,
each of thickness $d$, and
the result of summing multiple reflections is
\be
\mathcal{R}=\frac1{2R}\bigg[e^{2\kappa a}+R^2-T^2-\sqrt{\left(e^{2\kappa a}
-R^2-T^2\right)^2-4 R^2T^2}\bigg].
\ee
If the potentials consist of dielectric slabs, with dielectric constant
$\varepsilon$ and thickness $d$, the TE reflection and transmission
coefficients for a single slab are ($\kappa'=\sqrt{\varepsilon\zeta^2+k^2}$)
\begin{subequations}
\bea
R^{\rm TE}&=&\frac{e^{2\kappa' d}-1}{\left(\frac{1+\kappa'/\kappa}{1-\kappa'
/\kappa}\right)e^{2\kappa'd}-\left(\frac{1-\kappa'/\kappa}{1+\kappa'/\kappa}
\right)},\\
T^{\rm TE}&=&\frac{4(\kappa'/\kappa) e^{\kappa' d}}
{(1+\kappa'/\kappa)^2e^{2\kappa'd}-(1-\kappa'/\kappa)^2}.
\eea
\end{subequations}
The TM reflection and transmission coefficients are obtained by replacing,
 except in the exponents,
$\kappa'\to\kappa'/\varepsilon$.  (Multilayer potentials have been
discussed extensively in the past, see, for example, Refs.~\cite{zhou,tomas,
casbook09,parsegian}.)

\subsection{Casimir-Polder force}
\begin{figure}
\centering
\epsfig{file=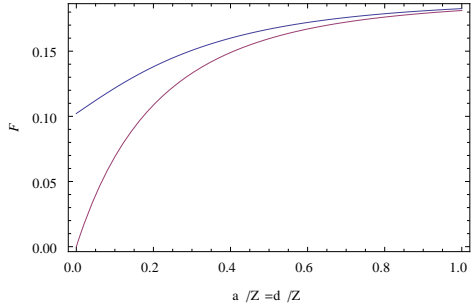,height=4cm}
\caption{\label{cpfig} Casimir-Polder energy between an semi-infinite array
of dielectric slabs with $\varepsilon=2$, compared to the energy (lower
curve) if only one slab were present.  Here we have assumed that the
spacing between the slabs and the widths of the slabs are equal.}
\end{figure}

\begin{figure}
\centering
\epsfig{file=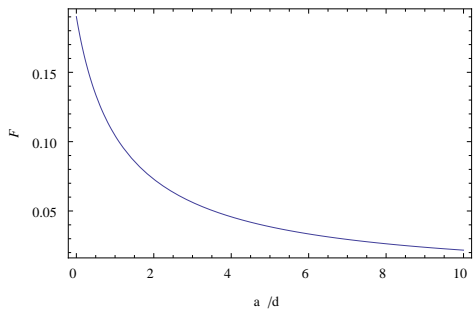,height=4cm}
\caption{\label{ztoinfty} Casimir-Polder energy for large distances from
the array, as a function of the ratio $a/d$, where $a$ is the
distance between the dielectric slabs in the array, and $d$ is
the thickness of each slab.  Here $\varepsilon=2$.}
\end{figure}

Consider an atom, of polarizability $\alpha(\omega)$, a distance $Z$ to the
left of the array.  The Casimir-Polder energy is
\be
E=-\int_{-\infty}^\infty
 d\zeta\int\frac{d^2k}{(2\pi)^2}\alpha(i\zeta)\mbox{tr}\,{\bf g}(Z,Z),
\ee
where apart from an irrelevant constant the trace is
\be \mbox{tr}\,{\bf g}(Z,Z)\to
\frac1{2\kappa}\left[-\zeta^2\mathcal{R}^{\rm TE}+(\zeta^2+2k^2)\mathcal{R}^{
\rm TM}\right]e^{-2\kappa|Z|}.
\ee
For example, in the static limit, where we disregard the frequency
dependence of the polarizability,
\be
E=-\frac{\alpha(0)}{2\pi}\frac1{Z^4}F(a/Z,d/Z).
\ee
This is compared with the single slab result in Fig.~\ref{cpfig}.
It is interesting to consider the 
$Z\to\infty$ limit, which is shown in Fig.~\ref{ztoinfty}.
When $a/d\to0$ we recover the bulk limit.

\section{Annular pistons}
\begin{figure}
\centering
\epsfig{file=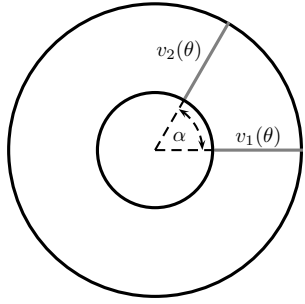,height=4cm}
\caption{\label{annpist} Two concentric Dirichlet cylinders, containing
semitransparent radial planes separated by an angle $\alpha$.
The radial planes constitute an annular piston.}
\end{figure}

The multiple scattering approach allows us to calculate the torque
between annular pistons, as illustrated in Fig.~\ref{annpist}.
We use multiple scattering in the angular coordinates,
and an eigenvalue condition in the radial coordinates---the problem is
equally well solvable
 with radial Green's functions, but 
this method permits interesting generalizations.
 
\begin{figure}
\centering
\epsfig{file=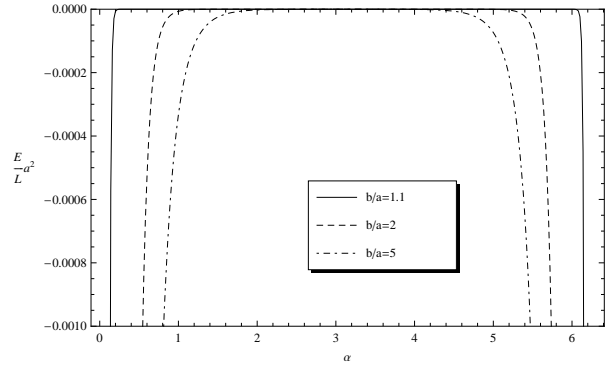,height=5cm}
\caption{\label{figanp} Energy of annular piston as a function
of the angle between Dirichlet planes $\lambda_1=\lambda_2=\infty$.}
\end{figure}

Using the argument principle to determine the angular eigenvalues,
we get the following expression for the energy 
between radial Dirichlet planes for an annular Casimir
piston: 
\bea\mathcal{E}=\frac{E}{L_z}&=&\frac1{8\pi^2 i}\int_0^\infty {\kappa d \kappa}
  \int\limits_\gamma d \eta
  \frac\partial{\partial\eta} \ln \left[
    K_{i\eta}(\kappa a)I_{i\eta}(\kappa b) -
    I_{i\eta}(\kappa a) K_{i\eta}(\kappa b)\right]\nn\\
 && \times\ln \left( 1- \frac{\lambda_1 \lambda_2 \cosh^2 \eta (\pi-\alpha)
/\cosh^2\eta\pi}
  {\left(2\eta \tanh \eta \pi +\lambda_1\right)
    \left(2 \eta \tanh \eta \pi +\lambda_2  \right) } \right),
\eea
where $I_{i\eta}$ and $K_{i\eta}$ are modified Bessel functions
of imaginary order, and the contour $\gamma$ encircles the poles in $\eta$
along the positive real axis.
As to be described elsewhere \cite{wedge3}, 
we have extracted numerical results for
the energy, as shown in Fig.~\ref{figanp}.

\begin{figure}
\centering
\epsfig{file=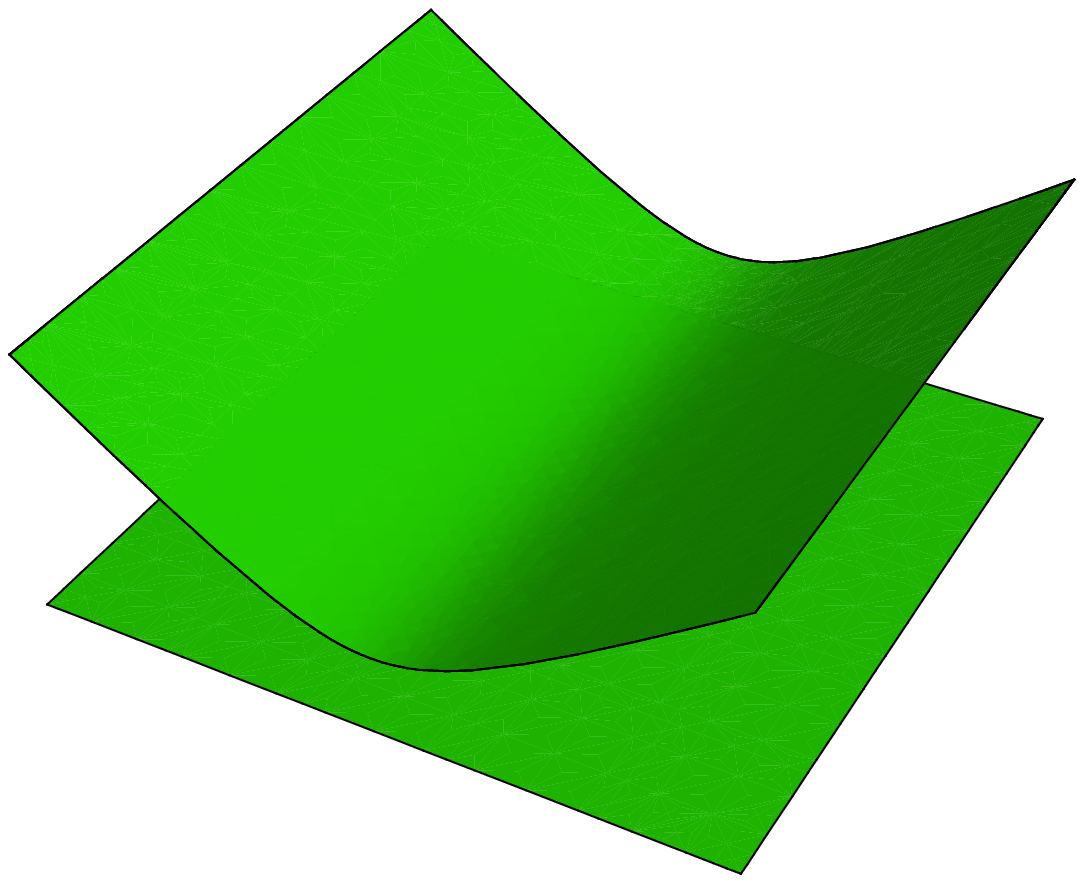,height=3cm}
\caption{\label{hyperpl} Hyperbolic cylinder above a plane.}
\epsfig{file=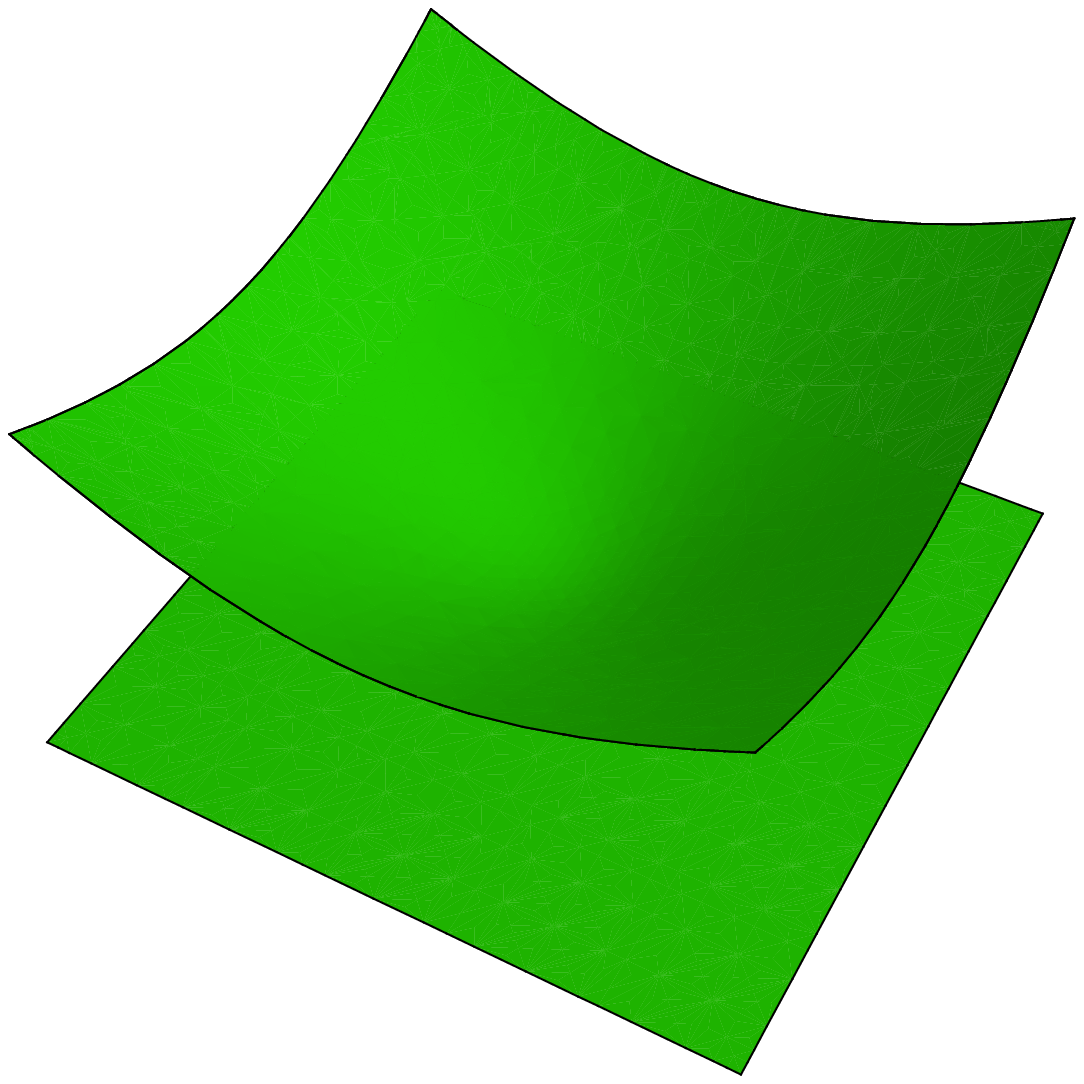,height=3cm}
\caption{\label{hyperrev} Hyperboloid of revolution above a plane.}
\end{figure}

We hope to apply these methods to study interactions between hyperbol\ae,
such as a hyperbolic cylinder above a plane, 
and a hyperbola of revolution above a plane, see Figs.~\ref{hyperpl}
and \ref{hyperrev}.

\section{Conclusions}
This survey describes some of the Oklahoma group's work applying
multiple scattering techniques to interesting geometries.  
We have extensively explored weak coupling, because such cases can be
carried out exactly, and therefore they serve as a laboratory for testing 
general features, such as edge effects and the range of validity of
the proximity force approximation.
We also have general results for the Green's functions for
arbitrary piecewise continuous potentials in separable coordinates.
From these we can calculate not only Casimir-Polder forces, but Casimir
energies and torques for many geometries, including  annular pistons,
and  forces between hyperbolic surfaces.
Finally, new results for electromagnetic non-contact gears,
both for conductors and dielectrics, are in progress.

\acknowledgments

We thank W.-j. Kim, U. Schwarz, A. Sushkov, and
S. Lamoreaux for organizing such a stimulating Workshop on Casimir
Forces and Their Measurement.  We also acknowledge financial support
from the National Science Foundation and the Department of Energy.
We thank I. Brevik, S. Ellingsen, and
 K. Kirsten for collaboration on the annular problem.  We especially
thank K.V. Shajesh for his contributions to the noncontact gears
calculations and the discussion of temperature effects.

\end{document}